\documentclass[10pt,twocolumn,aps,english,pra,superscriptaddress,showpacs]{revtex4}
\usepackage{amsfonts}
\usepackage[T1]{fontenc}
\usepackage[latin9]{inputenc}
\usepackage{amsmath}
\usepackage{amssymb}
\usepackage{graphicx}
\usepackage{babel}
\usepackage{booktabs}
\usepackage{tabu}
\usepackage{subfigure}
\usepackage{physics}
\usepackage{graphicx}
\usepackage{mathrsfs}
\usepackage{color}

\makeatletter
\@ifundefined{textcolor}{}
{
 \definecolor{BLACK}{gray}{0}
 \definecolor{WHITE}{gray}{1}
 \definecolor{RED}{rgb}{1,0,0}
 \definecolor{GREEN}{rgb}{0,1,0}
 \definecolor{BLUE}{rgb}{0,0,1}
 \definecolor{CYAN}{cmyk}{1,0,0,0}
 \definecolor{MAGENTA}{cmyk}{0,1,0,0}
 \definecolor{YELLOW}{cmyk}{0,0,1,0}
 }
\makeatother

\begin{document}

\title{Role of symmetry in quantum search via continuous-time quantum walk}

\author{Yunkai Wang}
\email{yunkaiw2@illinois.edu}
\affiliation{Department of Physics, University of Illinois at Urbana-Champaign, Illinois 61801, USA}

\author{Shengjun Wu}
\email{sjwu@nju.edu.cn}
\affiliation{Institute for Brain Sciences and Kuang Yaming Honors School, Nanjing University, Nanjing 210023,
China}

\date{\today }

\begin{abstract}

For quantum search via the continuous-time quantum walk, the evolution of the whole system is usually limited in a small subspace.   In this paper, we discuss how the symmetries of the graphs are related to the existence of such an invariant subspace, which also suggests a dimensionality reduction method based on group representation theory. We observe that in the one-dimensional subspace spanned by each desired basis state which assembles the identically evolving original basis states, we always get a trivial representation of the symmetry group. So we could find the desired basis  by exploiting the projection operator of the trivial representation. Besides being technical guidance in this type of problem, this discussion also suggests that all the symmetries are used up in the invariant subspace and the asymmetric part of the Hamiltonian is very important for the purpose of quantum search.

\end{abstract}

\maketitle

\section{Introduction}

Quantum walk has been widely discussed in several aspects due to its tractable theoretical features and friendly experimental implementation \cite{venegas2012quantum,kempe2003quantum}. Continuous-time and discrete-time quantum walk are the two popular types. Discrete-time quantum walk is described by the application of a series of unitary transformations which are defined with an extra coin Hilbert space. The continuous-time quantum walk (CTQW) is described by the continuous evolution governed by a Hamiltonian and the coin space is not introduced in the model. Just as for the classical random walk,  the basic dynamic feature, such as the hitting time  \cite{krovi2006quantum,krovi2007quantum}, and several possible applications have been discussed.  Quantum walk is used as tools to design the  quantum algorithms \cite{childs2003exponential,shenvi2003quantum,di2011mimicking} and implement the universal quantum computing \cite{childs2009universal,lovett2010universal,childs2013universal}. The mathematical structure, similar to lattice model in condensed matter, also makes it possible to use quantum walk to simulate topological phenomena \cite{flurin2017observing,xuenp2017}.

Grover's quantum search algorithm is one of the first well-known quantum algorithms, which aims at searching for a marked item and provides a square root speed up on an unstructured database \cite{PhysRevLett.79.325}. However, for a practical database, the structure of the database might put constraints on  a quantum search algorithm. The search on a structured database can be solved using CTQW  \cite{PhysRevA.70.022314}, which is equivalent to searching on a graph. They discussed the search on complete graph, hypercube and $d$-dimensional periodic lattice. Since then, quantum search via CTQW has been widely discussed on different graphs: strongly regular graphs \cite{PhysRevLett.112.210502},  truncated $M$-simplex lattice \cite{PhysRevLett.114.110503,wang2020optimal}, balanced trees \cite{PhysRevA.93.032305}, Erd\"{o}s-Renyi random graphs \cite{PhysRevLett.116.100501}, complete bipartite graphs, star graph \cite{novo2015systematic}, Johnson graphs \cite{wong2016quantum}, dual Sierpinski gasket, T fractal, Cayley trees \cite{agliari2010quantum,wang2019controlled}. Despite the intensive study of quantum search on many graphs, the condition to achieve an optimal runtime remains elusive. Some works have tried to clarify the properties of this algorithm. Connectivity of the graph is shown to be a poor indicator for faster search \cite{PhysRevLett.114.110503}. The global symmetry of the graph is shown to be unnecessary for achieving optimal runtime \cite{PhysRevLett.112.210502}.

In most of these existing discussions, the analysis is  conducted in a small invariant subspace of the Hamiltonian. The invariant subspace is usually spanned by the basis states, which are the equal superposition of each types of identically evolving original basis states. This problem of dimensionality reduction has been discussed using Lanczos algorithm \cite{novo2015systematic}. However, in their work, the chosen basis states do not necessarily group the identically evolving basis states together. This might cause problems when we want to analyse flow of probability amplitude on the graph. One other way to reduce the dimensionality is to exploit the symmetries of Hamiltonian. This method has been explored for discrete-time quantum walk as a tool to discuss the hitting time  \cite{krovi2006quantum,krovi2007quantum}. We here exploit symmetries for quantum search using CTQW. The mathematical tools used in the discussion is slightly different. More importantly,  the Hamiltonian does not respect all the symmetries of the graph due to the introduction of oracle in the quantum search. Besides  serving as guidence for dimensionality reduction, our discussion deepens our understanding of role played by symmetries in quantum search. In  the previous discussion about global symmetry of graph \cite{PhysRevLett.112.210502}, they construct a graph which is not globally symmetric and still supports optimal runtime. We here show the symmetries are all used up in the dimensionality reduction and the asymmetric part of the Hamiltonian will determine the behavior of the algorithm.

This structure of this paper is as follows. We introduce the continuous-time quantum walk and quantum search via CTQW in Sec. II. In Sec. III, we discuss the derivation of symmetries group for the Hamiltonian. In Sec. IV, we clarify the role of symmetries in the dimension reduction. We conclude in Sec. V.

\section{Continuous-time quantum walk via quantum search}

A continuous-variable quantum walk is usually defined on a graph $G=(V,E)$ \cite{farhi1998quantum}, where $V$ is the set of vertices and $E$ is  the set of edges.  The $i$th vertex is assigned basis state $\ket{i}$ and we will use the descriptions the basis state $\ket{i}$ and vertex $i$ interchangeably. The evolution of the system can be described as $i\frac{d}{dt}q_{j}(t)=\sum_{k\in V}L_{jk}q_{k}(t)$, where $q_{j}$ is the probability amplitude at vertex $j$, $L_{jk}$ is the Laplacian defined as
\begin{equation}
 L_{j,k}=\begin{cases}
-\text{deg}(j), & \quad j=k\\
1, & \quad(j,k)\in E\\
0, & \quad\text{otherwise},
\end{cases}
\end{equation}
where $\text{deg}(j)$ is the degree of vertex $j$, which is the weight sum of all edges connected to vertex $j$. This means the Hamiltonian of the CTQW is chosen as $\hat{H}=L$ if described with the basis $\{\ket{i}\}$. In the quantum search problem, we need to make the marked vertex special and hence we need to slightly modify the Hamiltonian. One way to do this is to introduce an oracle for the marked vertex $\ket{w}\bra{w}$ as in Ref. \cite{PhysRevA.70.022314},
\begin{equation}
\hat{H}=-\gamma L-\ket{w}\bra{w},
\end{equation}
where $\gamma$ is jumping rate, i.e. the probability per unit time of jumping to an adjacent vertex.

The simplest example of quantum search via CTQW is given on complete graph in Ref. \cite{PhysRevA.70.022314}. Complete graph is a regular graph, i.e. deg(j) is independent of j. So, all the diagonal elements of its Laplacian are the same, which can be dropped  by rezeroing the energy without affecting the analysis of the evolution of the state.  In this case, $\hat{H}=-\gamma A-\ket{w}\bra{w}$
 , where
\begin{equation}
A_{j,k}=\begin{cases}
\begin{array}{ccc}
1, &  & (j,k)\in E\\
0, &  & (j,k)\not\in E,
\end{array}\end{cases}
\end{equation}
is the adjacent matrix of a graph. In the implementation of CTQW using an array of waveguides on a photonic chip, the $i$th diagonal elements of Laplacian are determined by the propagating constant of the $i$th waveguide \cite{tang2018implementation}. Replacing Laplacian $L$ with adjacent matrix $A$ corresponds to changing the propagating constant of all waveguides by the same amount. The Hamiltonian of a quantum search on a complete graph is then
\begin{displaymath}
H =-\gamma\left[\begin{matrix}
1/\gamma & 1 & 1&\cdots\ &1\\ 1 & 0 &1 & \cdots\ & 1\\ 1 & 1 & 0&\cdots\ &1\\
\vdots&\vdots & \vdots & \ddots & \vdots \\ 1 & 1 & 1&\cdots\ & 0\\
\end{matrix} \right].
\end{displaymath}

\begin{figure}[bt]
\includegraphics[width=0.2\textwidth]{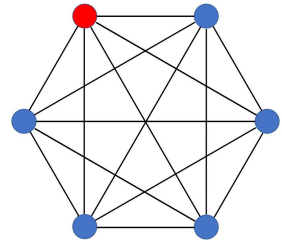}
\caption{Complete graph with 6 vertices. The red vertex $\ket{w}$ is the marked one. The blue vertices evolve indentically.}
\label{complete_graph}
\end{figure}

For a complete graph with $N$ vertices, the dimension of the Hilbert space is $N$. We would hope to analyse this problem in one of its invariant subspace. As shown in Fig. \ref{complete_graph}, the red vertex $\ket{w}$ is the marked one. All the other blue vertices evolve identically and we group them together to form a new basis state $\ket{b}=\frac{1}{\sqrt{N-1}}\sum_{i\not=w}\ket{i}$. In the subspace spanned by $\ket{w}$ and $\ket{b}$, the Hamiltonian is
\begin{displaymath}
H =-\gamma\begin{bmatrix} 1/\gamma&\sqrt{N-1}\\ \sqrt{N-1}&N-2\\
\end{bmatrix}.
\end{displaymath}


Since we have no information about the marked vertex, the initial state is chosen as the equal superposition of all vertices $\ket{s}=\frac{1}{\sqrt{N}}\sum\ket{i}$. We now want to solve for the evolution of the system $\ket{\psi(t)}=e^{-iHt}\ket{s}$. The goal is to tune the jumping rate $\gamma$ such that the probability amplitude can be concentrated on the marked vertex after a period of time of evolution. For the two-by-two matrix of the Hamiltonian described in the invariant subspace, we can easily find  when $\gamma N=1$ and $N$ is large enough, the eigenstates and the corresponding energies are $\ket{\psi_{0,1}}\approx(\ket{b}\pm\ket{w})/\sqrt{2}$, $E_{0,1}=-1\mp1/\sqrt{N}$. So, we can find the evolution of the system,
\begin{equation}
\ket{\psi(t)}=e^{-iHt}\ket{s}\approx\frac{1}{\sqrt{2}}(e^{-iE_{0}t}\ket{\psi_{0}}+e^{-iE_{1}t}\ket{\psi_{1}}).
\end{equation}
To find the marked vertex, we just need to do a projective measurement onto all basis  states of single vertex $\{\ket{i}\}_i$. This measurement will project the state onto the marked vertex $\ket{w}$ with success probability
{\begin{equation}
|\bra{w}\ket{\psi(t)}|^{2}=\frac{1}{2}[1-\cos(E_{0}-E_{1})t],
\end{equation}
 hence we could find the marked vertex $\ket{w}$ at time $t=\pi N^{1/2}/2$ with success probability close to $100\%$.} We can see that in the above analysis, it is very important to reduce the dimension of the Hilbert space by finding an invariant subspace. And this essentially comes from the symmetries of the graph. We will introduce more examples where this relation becomes more important.

\begin{figure}[bt]
\includegraphics[width=0.5\textwidth]{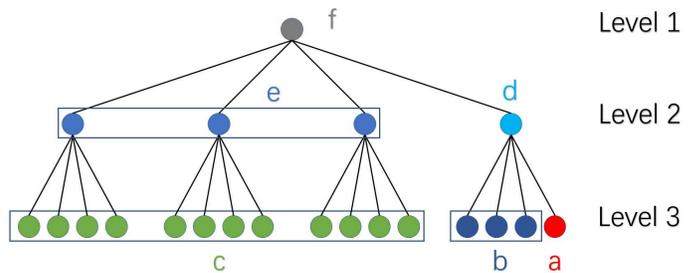}
\caption{A balanced tree with height $r=2$ and branching factor $M=4$. The invariant subspace is spanned by $\ket{a},\cdots,\ket{f}$ which are the equal superposition of vertices evolving identically as labeled by different colors. }
\label{tree}
\end{figure}

Consider the balanced tree with height $r$ and branching factor $M$, which is the number of vertices of the lower level connected to each vertex of the upper level. An example of balanced tree with $r=2$, $M=4$ is shown in  Fig.~\ref{tree}. {Notice balanced trees are not regular graphs, whose Laplacian won't have equal diagonal elements. So, we will use Laplacian $L$ instead of adjacent matrix $A$ in the Hamiltonian. } The quantum search via CTQW on a balanced tree has been studied in Ref. \cite{wang2019controlled}.  For a balanced tree with $N=(M^{r+1}-1)/(M-1)$ vertices, the dimension of the Hilbert space is again $N$. We would hope to analyse this problem in its invariant subspace to simplify the problem.  The evolution of the vertices labeled in the same color in Fig.~\ref{tree} is identical. And the evolution of the whole system is limited in an invariant subspace spanned by the following $6$ basis states:
$\ket{a}$ (the marked vertex state), $\ket{b}=\frac{1}{\sqrt{M-1}}\sum_{i\in b} \ket{i}$, $\ket{c}=\frac{1}{\sqrt{M(M-1)}}\sum_{i\in c} \ket{i}$, $\ket{d}$, $\ket{e}=\frac{1}{\sqrt{M-1}}\sum_{i\in e} \ket{i}$, and $\ket{f}$. In this subspace, the Hamiltonian can be written as the follow matrix,
{\small
\begin{displaymath}
H
=-\gamma\begin{bmatrix}
-1+\frac{1}{\gamma}& 0 & 0 & 1 & 0 & 0\\
0 & -1 & 0 & \sqrt{M\!-\!1} & 0 & 0\\
0 & 0 & -1 & 0 & \sqrt{M} & 0\\
1 & \sqrt{M\!-\!1} & 0 & -M\!-\!1 & 0 & 1\\
0 & 0 & \sqrt{M} & 0 & -M\!-\!1 & \sqrt{M\!-\!1}\\
0 & 0 & 0 & 1 & \sqrt{M\!-\!1} & -M\\
\end{bmatrix} .
\end{displaymath}
}
To look for the success probability, we again need to find its spectrum and calculate the evolution starting from the initial state  $\ket{s}=\frac{1}{\sqrt{N}}\sum\ket{i}$.  It turns out that we will need a two-stage search process. When $\gamma=2$, $\ket{\psi}_{0,1}\approx(\ket{b}\pm\ket{s})/\sqrt{2}$ with $E_1-E_0\approx 4 M^{-3/2}$ and
\begin{equation}
|\bra{b}\ket{\psi(t)}|^2=[1-\cos (E_1-E_0)t]/2.
\end{equation}
This means  the probability amplitude will be shifted to $\ket{b}$ after time $T_1=\pi M^{3/2}/4$ in the first stage, which is the runtime of the first stage of the algorithm. And for the second stage, it is found that when $\gamma=1$, $\ket{\psi}_{0,2}\approx(\ket{b}\pm\ket{a})/\sqrt{2}$ with $E_2-E_0\approx 2 M^{-1/2}$ and
\begin{equation}
|\bra{a}\ket{\psi(t)}|^2=[1-\cos (E_2-E_0)t]/2.
\end{equation}
So, the second stage takes time $T_2=\pi M^{1/2}/2$, we can find the marked vertex by projective measurement onto all the basises $\{\ket{i}\}$. And with success probabilty close to $100\%$, the outcome we get is the marked vertex.

\begin{figure}[bt]
\includegraphics[width=0.5\textwidth]{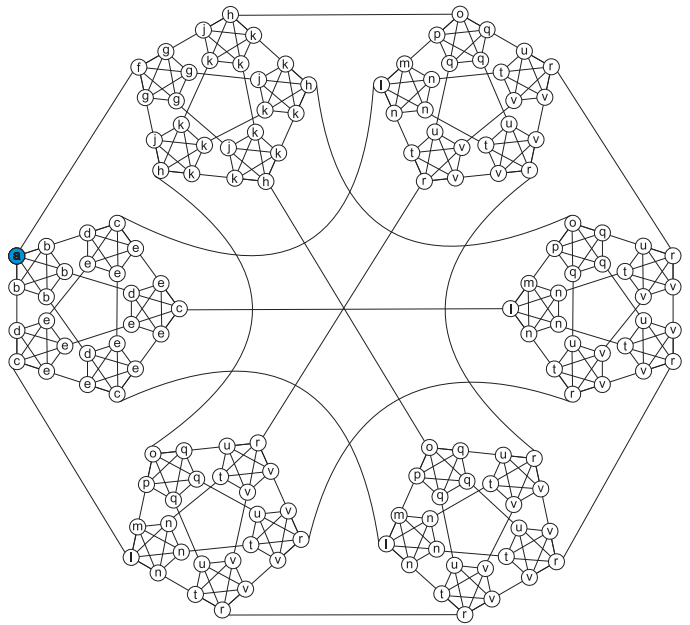}
\caption{Second-order truncated five-simplex lattice with identically evolving vertices labeled with the same letter. The blue vertex is the marked vertex $a$.}
\label{simplex}
\end{figure}

A more involved example is the 2nd-order truncated simplex lattice and the quantum walk via CTQW on it has been studied in Ref. \cite{wang2020optimal}. The definition of a truncated $M$-simplex lattice starts from a complete graph with $(M+1)$ vertices, which is the zeroth order truncated $M$-simplex lattice. To get the $(j+1)$th order truncated $M$-simplex lattice, we replace every vertex with a complete graph with $M$ vertices.  More details about the definition of a truncated $M$-simplex lattice can be found in Ref. \cite{J.Math.Phys.18.577}.  An example of 2nd-order truncated 5-simplex lattice is shown in Fig.~\ref{simplex}. {Notice truncated simplex lattices are also regular graphs and we can use adjacent matrix $A$ in the Hamiltonian by rezeroing the energy. But we will use Laplacian $L$ in the analysis of the symmetry as the evolution of the system with Hamiltonian using $A$ and $L$ will be the same.} The marked vertex is on the outer layer and labeled in blue. To analyse the evolution of the system, we need to find the Hamiltonian in an invariant subspace, which has been given in Ref. \cite{wang2020optimal}. 
We can then again find its spectrum and it turns out that quantum search on this graph needs three stages. The jumping rate is chosen as $\gamma=3/M,2/M,1/M$ respectively, with runtime $T_{1,2,3}=\pi M^{5/2}/6, \pi M^{3/2}/4, \pi M^{1/2}/2$ in each stage. And success probability close to $100\%$ can be achieved.

In the process of analysing quantum search on the above examples, an important step is to reduce the dimensionality of the Hilbert space so that we could find the spectrum of the Hamiltonian, which is an important starting point of the analysis. For complete graphs, the structure of the graph is so simple that we could group the identically evolving vertices intuitively. This task obviously becomes more involved for the other examples. We will discuss the relation between this dimension reduction and the symmetry of the graphs. This can provide some guidance to the reduction calculation. And more importantly, this discussion can give us hint about the role played by symmetry in quantum search.


\section{Symmetry group of the Hamiltonian}

Most of the discussions about quantum search via CTQW were conducted on the graphs with some symmetries. Although it has been argued that the global symmetry is unnecessary for the search algorithm \cite{PhysRevLett.112.210502}, the graphs they chose also contain some symmetries. Otherwise, we need to discuss the problem in a very large Hilbert space which will make it hard to analyse the proper value of $\gamma$ and the runtime of the algorithm. We will discuss the symmetries of Hamiltonian partially inherited from the graphs in this section.

Let's first discuss what is the mathematical description of symmetries of graphs. If $g$ is a symmetry of a graph and acts like a permutation of vertices, applying $g$ should not affect the Laplacian $L$ of the graph. In other word, we would have $D(g)LD(g^{-1})=L$, where $D(g)=T_{i_1j_1}T_{i_2j_2}\cdots T_{i_mj_m}$ is the matrix representation of $g$. The matrix $T_{i_1j_1}$ corresponds to the permutation of vertex $i_1,j_1$ and is the matrix given by exchanging the row $i$ and row $j$ of the identity matrix.

In a quantum mechanics system with symmetry $G=\{g_i\}$, the Hilbert space is the representation space for $G$, in which each $g_i$ will act like a matrix. If $g_i$ is a symmetry of the system, then $g_i$ should commute with Hamiltonian, i.e. $[g_i,\hat{H}]=0$. Without the oracle, the Hamiltonian $H=L$ should inherit all the symmetries from the graph. And each vertex $i$ corresponds to a basis $\ket{i}$ of the Hilbert space. Consider the symmetries of $\hat{H}=L$  first and take the complete graph as an example, for which its symmetries consist of all the permutation of $\{\ket{1},\ket{2},\cdots, \ket{N}\}$, i.e. $S_N$. In the Hilbert space, the representation of each permutation could again be written as the product of a series of row-switching elementary matrix, i.e. $D(g)=T_{i_1j_1}T_{i_2j_2}\cdots T_{i_mj_m}$, where $T_{i_1j_1}$ corresponds to the permutation of basis $\ket{i_1},\ket{j_1}$ and is the matrix given by exchanging the row $i_1$ and row $j_1$ of the identity matrix. Then $D(g^{-1})=T_{i_mj_m}^{-1}T_{i_{m-1}j_{m-1}}^{-1}\cdots T_{i_1j_1}^{-1}=T_{i_mj_m}T_{i_{m-1}j_{m-1}}\cdots T_{i_1j_1}$. $g$ is a symmetry of the Hamiltonian if $D(g^{-1})HD(g)=H$. It is obvious that when the graph has some symmetries $\{g\}$ of switching vertices, then we always have $D(g^{-1})HD(g)=H$, which can be checked by doing the matrix multiplication using the Laplacian.

\begin{figure}[!pbt]
\includegraphics[width=0.5\textwidth]{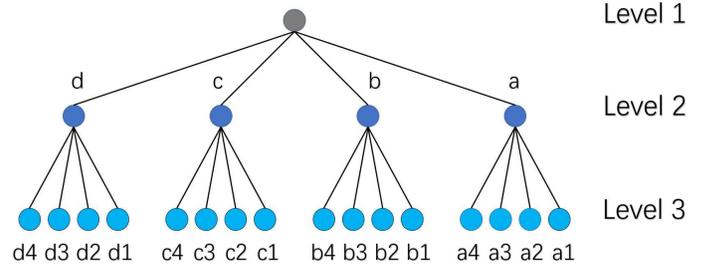}
\caption{Balanced tree with height$=2$ and branching factor $M=4$. The marked vertex is vertex $a_1$.}
\label{cayley_tree}
\end{figure}

We then consider a non-trivial example, a balanced tree as shown in Fig. \ref{cayley_tree}. The symmetries of this graph  include: (1)All of the permutation of vertices in each branch. For example, all the permutation of $\{a1,a2,a3,a4\}$. (2)Exchange the position of the whole branch. For example, $a\rightarrow b$, $a1\rightarrow b1$, $a2\rightarrow b2$, $a3\rightarrow b3$, $a4\rightarrow b4$. The Laplacian of balanced tree of height$=2$ and branching factor $M=4$ is given in Table \ref{L_Cayley_tree}. For the above symmetry $g$ of the graph,  we can check  $D(g^{-1})LD(g)=L$ is always true.

\begin{table*}[tb]
\normalsize
\centering
\caption{The Laplacian $L$ of balanced tree of height$=2$ and branching factor $M=4$. \label{L_Cayley_tree}}
\begin{tabular}{|c|c|c|c|c|c|c|c|c|c|c|c|c|c|c|c|c|c|c|c|c|c|}
\hline
$L$ & $a_1$ & $a_2$ & $a_3$ & $a_4$ & $b_1$ & $b_2$ & $b_3$ & $b_4$ & $c_1$ & $c_2$ & $c_3$ & $c_4$ & $d_1$ & $d_2$ & $d_3$ & $d_4$ & $a$ & $b$ & $c$ & $d$ & $e$\\
\hline
$a_1$ & -1 & 0 & 0 & 0 & 0 & 0 & 0 & 0 & 0 & 0 & 0 & 0 & 0 & 0 & 0 & 0 & 1 & 0 & 0 & 0 & 0\\
\hline
$a_2$ & 0 & -1 & 0 & 0 & 0 & 0 & 0 & 0 & 0 & 0 & 0 & 0 & 0 & 0 & 0 & 0 & 1 & 0 & 0 & 0 & 0\\
\hline
$a_3$ &  0 & 0 & -1 & 0 & 0 & 0 & 0 & 0 & 0 & 0 & 0 & 0 & 0 & 0 & 0 & 0 & 1 & 0 & 0 & 0 & 0\\
\hline
$a_4$ &  0 & 0 & 0 & -1 & 0 & 0 & 0 & 0 & 0 & 0 & 0 & 0 & 0 & 0 & 0 & 0 & 1 & 0 & 0 & 0 & 0\\
\hline
$b_1$ &  0 & 0 & 0 & 0 & -1 & 0 & 0 & 0 & 0 & 0 & 0 & 0 & 0 & 0 & 0 & 0 & 0 & 1 & 0 & 0 & 0\\
\hline
$b_2$ &  0 & 0 & 0 & 0 & 0 & -1 & 0 & 0 & 0 & 0 & 0 & 0 & 0 & 0 & 0 & 0 & 0 & 1 & 0 & 0 & 0\\
\hline
$b_3$ &  0 & 0 & 0 & 0 & 0 & 0 & -1 & 0 & 0 & 0 & 0 & 0 & 0 & 0 & 0 & 0 & 0 & 1 & 0 & 0 & 0\\
\hline
$b_4$ &  0 & 0 & 0 & 0 & 0 & 0 & 0 & -1 & 0 & 0 & 0 & 0 & 0 & 0 & 0 & 0 & 0 & 1 & 0 & 0 & 0\\
\hline
$c_1$ &  0 & 0 & 0 & 0 & 0 & 0 & 0 & 0 & -1 & 0 & 0 & 0 & 0 & 0 & 0 & 0 & 0 & 0 & 1 & 0 & 0\\
\hline
$c_2$ &  0 & 0 & 0 & 0 & 0 & 0 & 0 & 0 & 0 & -1 & 0 & 0 & 0 & 0 & 0 & 0 & 0 & 0 & 1 & 0 & 0\\
\hline
$c_3$ &  0 & 0 & 0 & 0 & 0 & 0 & 0 & 0 & 0 & 0 & -1 & 0 & 0 & 0 & 0 & 0 & 0 & 0 & 1 & 0 & 0\\
\hline
$c_4$ &  0 & 0 & 0 & 0 & 0 & 0 & 0 & 0 & 0 & 0 & 0 & -1 & 0 & 0 & 0 & 0 & 0 & 0 & 1 & 0 & 0\\
\hline
$d_1$ &  0 & 0 & 0 & 0 & 0 & 0 & 0 & 0 & 0 & 0 & 0 & 0 & -1 & 0 & 0 & 0 & 0 & 0 & 0 & 1 & 0\\
\hline
$d_2$ &  0 & 0 & 0 & 0 & 0 & 0 & 0 & 0 & 0 & 0 & 0 & 0 & 0 & -1 & 0 & 0 & 0 & 0 & 0 & 1 & 0\\
\hline
$d_3$ &  0 & 0 & 0 & 0 & 0 & 0 & 0 & 0 & 0 & 0 & 0 & 0 & 0 & 0 & -1 & 0 & 0 & 0 & 0 & 1 & 0\\
\hline
$d_4$ &  0 & 0 & 0 & 0 & 0 & 0 & 0 & 0 & 0 & 0 & 0 & 0 & 0 & 0 & 0 & -1 & 0 & 0 & 0 & 1 & 0\\
\hline
$a$ & 1 & 1 & 1 & 1 & 0 & 0 & 0 & 0 & 0 & 0 & 0 & 0 & 0 & 0 & 0 & 0 & -5 & 0 & 0 & 0 & 1\\
\hline
$b$ & 0 & 0 & 0 &0 & 1 & 1 & 1 & 1 & 0 & 0 & 0 & 0 & 0 & 0 & 0 & 0 & 0 & -5 & 0 & 0 & 1\\
\hline
$c$ & 0 & 0 & 0 &0 & 0 & 0 & 0& 0 & 1 &1 & 1 & 1 & 0 & 0 & 0 & 0 & 0 & 0 & -5 & 0 & 1\\
\hline
$d$ & 0 & 0 & 0 &0 & 0 & 0 & 0& 0 & 0 &0 & 0 & 0 & 1 & 1 & 1 & 1 & 0 & 0 & 0 & -5 & 1\\
\hline
$e$ & 0 & 0 & 0 &0 & 0 & 0 & 0& 0 & 0 &0 & 0 & 0 & 0 & 0 & 0 & 0 & 1 & 1 &1 & 1 & -4\\
\hline
\end{tabular}
\end{table*}

\begin{figure}[!pbt]
\includegraphics[width=0.5\textwidth]{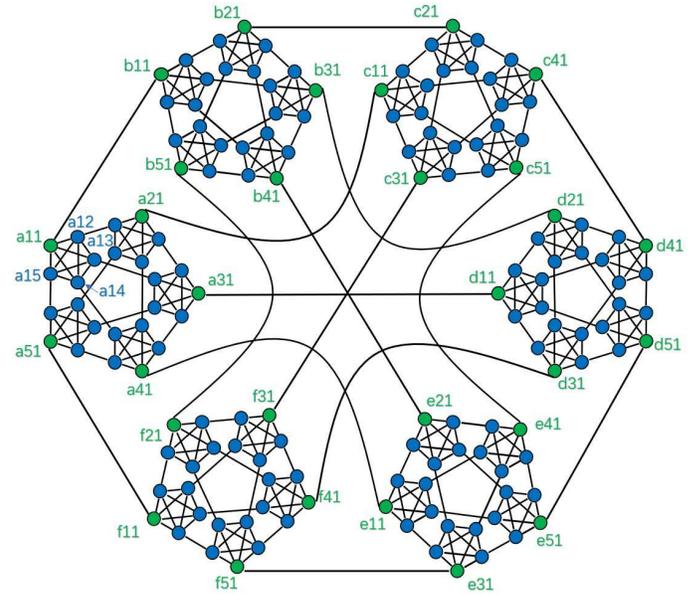}
\caption{2nd-order truncated 5-simplex lattice. The marked vertex is vertex $a11$.}
\label{simplex_group}
\end{figure}

And similarly, we can consider the symmetries of 2nd-order truncated simplex lattice. As shown in Fig.~\ref{simplex_group}, we label the vertices such that each large layer corresponds to a letter from $a,b,c,d,e,f$ and number further labels the position within the layer. The symmetries of 2nd-order truncated simplex lattice is that: The permutations,  $a{12}\leftrightarrow a{13}$, $a{2j}\leftrightarrow a{3j}$, $a{54}\leftrightarrow a{53}$, $a{45}\leftrightarrow a{44}$,   $b{12}\leftrightarrow b{13}$, $b{2j}\leftrightarrow b{3j}$, $b{54}\leftrightarrow b{53}$, $b{45}\leftrightarrow b{44}$, $e{14}\leftrightarrow e{15}$, $e{4j}\leftrightarrow e{5j}$, $e{21}\leftrightarrow e{25}$, $e{12}\leftrightarrow e{11}$, $f{13}\leftrightarrow f{14}$, $f{3j}\leftrightarrow f{4j}$, $f{15}\leftrightarrow f{14}$, $f{51}\leftrightarrow f{55}$, $c{ij}\leftrightarrow d{ij}$  for all $i,j$. We can do this permutation for any two green vertices connected to the same blue vertex without affecting the Laplacian.


One should notice that the oracle introduced for the marked vertex will break some of the symmetries of the original graph. When we consider the Hamiltonian for the quantum search via CTQW $H=-\gamma L-\ket{w}\bra{w}$, we need to consider a subgroup of the symmetries inherited from the graph. As argued in \cite{PhysRevLett.112.210502}, the global symmetry is unnecessary for the quantum search algorithm. Actually, we are introducing asymmetries with oracle to make the search process succeed. For the complete graph, when we introduce the oracle $-\ket{w}\bra{w}$ to the Hamiltonian, we break all the symmetries that exchange vertex $w$ with any other vertices.

For the balanced tree of height 2, when the marked vertex is one of the leaf vertex $a_1$ as shown in Fig. \ref{cayley_tree}, we break the symmetries of exchanging $a_1$ with all the other vertices $\{a_i\}$ in $a$ branch and the symmetries of exchanging the whole branch $a$ with any other branches. The actual symmetry group consists of all the permutations of vertices except vertex $a_1$ in branch $a$ and all permutations of whole branches except branch $a$. This can be derived by writing down all the symmetries of the original graphs and then delete all elements which move the position of the marked vertex. This process exactly gives us all the elements of the actual symmetry group. Because on the one hand, it is obvious that all the symmetries which move the marked vertex is no longer a symmetry. And the other symmetries remain a symmetry of the Hamiltonian because if we look at their action on the Laplacian, introducing an oracle to the Hamiltonian only affects one diagonal element, which is not toughed by these group elements which do not move the marked vertex. For 2nd-order truncated simplex lattice, when the marked vertex is a11 as shown in Fig.~\ref{simplex_group}, we also break all symmetries which move the position of a11.


\section{Group the identically evolving vertices exploiting projector}

Having understood the symmetries of Hamiltonian in quantum search via CTQW, we now consider how these symmetries give us a set of basis states which can span an invariant subspace for the evolution. We first classify all the vertices $\{\ket{1},\ket{2},\cdots,\ket{N}\}$ into small equivalent sets $V_1,V_2,\cdots$. For any two vertices $\ket{i},\ket{j}$, if there exists a symmetry $g$ such that $\ket{j}=g\ket{i}$, we assign them into the same set $V_\alpha$. We will show later that the vertices within each set will evolve identically. We emphasize that even if two vertices are in the same set, we cannot always expect to switch their positions without affecting the Laplacian of the graph. For example, switching $b_1$ and $c_1$ of Fig. \ref{cayley_tree} will  affect the Laplacian. But $b_1,c_1$ are in the same set by our definition because the symmetry $g$ of switching the whole branches of $b$ and $c$ will satisfy $\ket{c_1}=g\ket{b_1}$.

 We define the basis states $\{\ket{eq_\alpha}\}$ as the equal superposition of all basis states from the same set $V_\alpha$, i.e. $\ket{eq_\alpha}=\frac{1}{|V_\alpha|}\sum_{i\in V_\alpha}\ket{i}$. Because all the group elements only permute the vertices in the same set by definition, we have $g\ket{eq_\alpha}=\ket{eq_\alpha}$ for  $\forall\alpha, g$. We now prove $\{\ket{eq_i}\}$ form an invariant subspace of the evolution by contradiction. Assume there exists $e^{-iHt}\ket{eq_i}=span\{\ket{eq_i}\}+a\ket{i}$, where $a\neq 0$ is a constant, $\ket{i}$ does not belong to the subspace spanned by $\{\ket{eq_i}\}$, i.e. $i\notin span\{\ket{eq_i}\}$ . But $\ket{i}$ must belong to one of the equivalent sets $V_\alpha$ since they cover all the vertices. If $V_\alpha$ only has one element $\ket{i}$, then  $\ket{i}$ already belong to $span\{\ket{eq_i}\}$, which gives the contradiction. If $V_\alpha$ has more than one elements, we can pick another different element $\ket{j}\neq \ket{i}$. Since $\ket{i},\ket{j}$ belong to the same equivalent set, there must exist $g$ such that $\ket{j}=g\ket{i}$ by definition. We then apply $g$ to the evolution equation and find
\begin{equation}
\begin{aligned}
&ge^{-iHt}\ket{eq_i}=gspan\{\ket{eq_i}\}+ag\ket{i}\\
&\rightarrow ge^{-iHt}g^{-1}\ket{eq_i}=span\{\ket{eq_i}\}+a\ket{j}\\
&\rightarrow e^{-iHt}\ket{eq_i}=span\{\ket{eq_i}\}+a\ket{j},\\
\end{aligned}
\end{equation}
where in the second line we have used the fact that $g^{-1}\ket{eq_i}=\ket{eq_i}$ for $\forall\, i, g$, and in the third line, we have used the fact that $ge^{-iHt}g^{-1}=e^{-iHt}$ because $g$ is a symmetry of the Hamiltonian. Comparing this equation with the original evolution equation gives
\begin{equation}
\begin{aligned}
& e^{-iHt}\ket{eq_i}=span\{\ket{eq_i}\}+a\ket{j}\\
&e^{-iHt}\ket{eq_i}=span\{\ket{eq_i}\}+a\ket{i}\\
&\rightarrow \ket{i}=\ket{j},
\end{aligned}
\end{equation}
where $\ket{i}=\ket{j}$ contradicts with the fact that they are different vertices. We have thus proved  $\{\ket{eq_i}\}$ form an invariant subspace of the evolution.  Since the initial state $\ket{s}=\frac{1}{\sqrt{N}}\sum\ket{i}$ is within the subspace spanned by $\{\ket{eq_i}\}$, we could conclude the system will evolve within this subspace.

We now prove that the vertices in each set $V_\alpha$ will evolve identically. We emphasize in the following discussion we always choose the initial state as $\ket{s}=\frac{1}{\sqrt{N}}\sum\ket{i}$, i.e. the equal superposition of all original basis states.  If two vertices $\ket{i},\ket{j}$ are in the same equivalent set $V_\alpha$, then there exists  $g$ such that $\ket{j}=g\ket{i}$. Starting from initial state $\ket{s}$, the evolution of each vertex $i,j$ is described by the probability amplitude  $\psi_i(t)=\bra{i}e^{-iHt}\ket{s}$, $\psi_j(t)=\bra{j}e^{-iHt}\ket{s}$. Since $g$ is a symmetry of the Hamiltonian $H$, $e^{-i Ht}=ge^{-i Ht} g^{-1}$. We notice $g$ is just a permutation of basis states $\{\ket{i}\}$, then $g\ket{s}=\ket{s}$. $\psi_i(t)=\bra{i}e^{-iHt}\ket{s}=\bra{i}ge^{-iHt}g^{-1}\ket{s}=\bra{j}e^{-iHt}\ket{s}=\psi_j(t)$. We have thus proved the evolution of vertices $i,j$ are identical given there exists symmetry $g$ such that $[g,H]=0$ and $g\ket{i}=\ket{j}$. 



The group elements can be represented in the Hilbert space by matrices with a basis. If we rechoose the basis more carefully, the matrices of all group elements might be block diagonalized at the same time, we then find several subrepresentations. Sometimes there is no non-trivial subrepresentation. Then, they are irreducible representation, or \textit{irrep}. In each one-dimensional subspace spanned by $\{\ket{eq_i}\}$, we get an \textit{irrep} $A_1$ and the representation is simply a scalar $1$ for $\forall g\in S$. Each \textit{irrep} $A_1$ should be provided by the one-dimensional subspace which is spanned by a basis state $\{\ket{eq_i}\}$. No \textit{irrep} $A_1$ is spanned by basis states other than $\ket{eq_i}$ because for each subspace spanned by the set $V_\alpha=\{\ket{\alpha_1},\ket{\alpha_2},\cdots,\ket{\alpha_m}\}$, the only basis state $\ket{\psi}$ which satisfies $g\ket{\psi}=\ket{\psi}$ is $\ket{\psi}=\sum_i\ket{\alpha_i}/\sqrt{m}$.



To find the basis states $\{\ket{eq_i}\}$, we can use the projection operator onto representation space of irrep $A_1$. Projection operator is $P^J=\frac{dim J}{|G|}\sum_{g\in G}[\chi ^J (g)]^* g$, where $J$ labels \textit{irrep}, $dim J$ is the dimension of the \textit{irrep} $J$, $\chi^J(g)$ is the character of group element $g$ in the $irrep$ $J$. We want to find the basis state which gives trivial \textit{irrep} $A_1$. $P^{A_1}=\frac{dim A_1}{|S|}\sum_{g\in S}[\chi ^{A_1} (g)]^* g=\frac{1}{|S|}\sum_{g\in S} g$. Projection operator $P^{A_1}$ could project a basis state into the subspace $\mathscr{H}^{A_1}$. However, notice \textit{irrep} $A_1$ occurs $n$ times, $P^{A_1}$ will project to the reducible subspace $\mathscr{H}^{A_1}\oplus\mathscr{H}^{A_1}\oplus\cdots\oplus\mathscr{H}^{A_1}$, in which the Hamiltonian $\hat{H}$ acts as an $n\times n$ matrix.  We would like to point out that any linear combination of $\{\ket{eq_i}\}$ could span an one-dimensional subspace which supports an \textit{irrep} $A_1$. But when we apply projection operator $P^{A_1}$ to a single basis state $\ket{i}$, we would get one of the expected basis states $\{\ket{eq_i}\}$ rather than their combination. Because for $\forall g\in S$, $g\ket{i}$ always gives elements in the same equivalent set $V_\alpha$.



Based on the above observations, we are ready to find the expected basis states exploiting the projection operator $P^{A_1}$. The whole process is: (1)  Find symmetries of the graph. (2)  Delete the symmetries which has moved the marked vertex, which gives  the symmetries of Hamiltonian as we have discussed in the previous section. (3) Apply the projection operator $P^{A_1}$ to each of the original basis states $\ket{i}$. But whenever we find a $\ket{eq_i}$, we could skip all the vertices included in this $\ket{eq_i}$. Repeat applying the projector operation $P^{A_1}$ until we have assigned each $\ket{i}$ to one of $\ket{eq_i}$.

Take the complete graph as an example. Its symmetries of the Hamiltonian are all of the permutations of vertices except the marked one $w$, i.e. a permutation group $S_{N-1}$. If we apply $P^{A_1}$ to the marked vertex $\ket{w}$, we would get $\ket{w}$, since all elements in symmetry group of the Hamiltonian do nothing to the marked vertex. If we apply $P^{A_1}$ to one of the vertex $\ket{i}$ other than the marked one $\ket{i}\neq\ket{w}$, $P^{A_1}\ket{i}=\frac{1}{|S|}\sum_{g\in S} g \ket{i}=\frac{1}{(N-1)!}(N-2)!\sum_{i\not=w}\ket{i}\propto \ket{b}$ as we selected in the Sec. II based on intuition. Then we consider the balanced tree of height 2 as shown in Fig. \ref{cayley_tree}. If we introduce the oracle for the marked vertex $a_1$, the symmetries of the Hamiltonian have been found in Sec. III. Applying $P^{A_1}$ to $\ket{a_1}$ does nothing, $P^{A_1}\ket{a_1}\propto\ket{a_1}$. $\ket{a_1}$ is one of the $\{\ket{eq_i}\}$. $P^{A_1}\ket{a_2}\propto\sum_{i\in branch \, a}\ket{a_i}$. We hence skip all the other vertices in level 3 of branch $a$. $P^{A_1}\ket{a}\propto\ket{a}$. $P^{A_1}\ket{b_1}\propto \sum_{i\in \theta}\ket{i}$, where $\theta$ means the set of all vertices in the level 3 except the branch $a$. We then skip all the other vertices in level 3. $P^{A_1}\ket{b}\propto \sum_{i\in \alpha}\ket{i}$, where $\alpha$ means the set of all the vertices in the level 2 except $\ket{a}$. $P^{A_1}\ket{d}\propto \ket{d}$. Hence, we find the expected basis states $\{\ket{eq_i}\}$ are $\{\ket{a_1}, \ket{a}, \sum_{i\in branch \, a}\ket{a_i}, \sum_{i\in \theta}\ket{i},  \sum_{i\in \alpha}\ket{i}, \ket{d}\}$. After relabelling each basis state, we show them in Fig. \ref{tree}. In the subpsace spanned by these basis states, each matrix element of the Hamiltonian can be calculated, for example $\bra{a}H\ket{d}$ is one of the matrix element. Similar process can be done for the 2nd order truncated simplex lattice. Once the symmetries of Hamiltonian is derived, we can easily write down the basis states for an invariant subspace. The results are shown in Fig. \ref{simplex}, where each letter corresponds to a basis states $\ket{eq_i}$.

If we apply the dimensionality reduction method based on Lanczos algorithm proposed in \cite{novo2015systematic}, we would not get these expected basis states. In Fig. \ref{tree}, start from $\ket{a}$, $H\ket{a}\propto (-1+1/\gamma)\ket{a}+\ket{d}$, then one of the basis state is $\ket{d}$. But $H\ket{d}\propto \ket{a}+\sqrt{M-1}\ket{b}+(-M-1)\ket{d}+\ket{f}$, the third basis state given by their method would be $\sqrt{M-1}\ket{b}+\ket{f}$, which is not what we expect. Although this method could also give a set of basis states which span the same subspace as $\{\ket{a},\ket{b},\ket{c},\ket{d},\ket{e},\ket{f}\}$,  this way of choosing basis states might hide some physics within the algorithm. For example, we could not observe the flow of probability amplitude from $\ket{c}$ to $\ket{b}$, and from $\ket{b}$ to $\ket{a}$ as discussed in \cite{wang2019controlled}.

Notice that if we do not introduce the oracle to break the symmetries  in Fig. \ref{cayley_tree} , the basis states we get would be $P^{A_1}\ket{a_1}=\sum_{i\in layer\,3}\ket{i}$, $P^{A_1}\ket{a}=\sum_{i\in layer\,2}\ket{i}$, $P^{A_1}\ket{d}=\ket{d}$. The number of $\{\ket{eq_i}\}$ is actually smaller. This is reasonable because the more group elements there are in the symmetry group, the more complicated their relation should be. And then it is harder to find an invariant subspace and the dimension of each \textit{irrep} might increase. When we have more symmetries or equivalently the order of the symmetry group of the Hamiltonian $|G|$ increases, since $\sum_J dim\, J=N$, $\sum_J (dim\, J)^2=|G|$,  the number of \textit{irrep} should decrease, and the dimension of each \textit{irrep} might increase. Hence, we might also expect the number \textit{irrep} $A_1$ decreases. So, when the graph has more symmetries, we might find a  smaller matrix in the subspace for the Hamiltonian after reducing the dimensionality using symmetries.

\section{Conclusion and discussion}

We discussed the role of symmetries in quantum search via CTQW, especially by connecting it to the dimensionality reduction of the Hamiltonian. By observing the symmetries of the graph, we would find the symmetry group of the Laplacian $L$. After introducing the oracle into the Hamiltonian, we could find the symmetry group of the Hamiltonian by removing all elements which move the positions of the marked vertex. Then exploiting the projection operator $P^{A_1}$, we could find the expected basis states which is the equal superposition of identically evolving vertices $\{\ket{eq_i}\}$.

Since we have exploited all of the symmetries in the Hamiltonian, the Hamiltonian should not contain any other symmetries  in the subspace spanned by $\{\ket{eq_i}\}$. In some sense, we could conclude it is the asymmetry of the original graph and the asymmetry we introduced by the oracle term that play an important role in the success of the quantum search via CTQW. There exists the claim that global symmetry is unnecessary \cite{PhysRevLett.112.210502}. We would like to further suggest that asymmetry is necessary. This observation gives us a hint that asymmetry might play an important role just as the vital role they played in some other quantum information tasks \cite{gour2009measuring,hall2012does}.


\acknowledgments

This work is supported by the National Key R\&D Program of China (Grants No. 2017YFA0303703 and No. 2016YFA0301801)
and the National Natural Science Foundation of China (Grant No. 11475084).

\end{document}